\documentclass{nature}
\usepackage{bm}
\usepackage{hyperref}
\usepackage{amssymb}
\usepackage{epsfig}
\usepackage{color}
\usepackage{amssymb,amsmath}
\usepackage{aas_macros}
\usepackage{graphicx,epstopdf}

\newcommand{\be}{\begin{equation}}
\newcommand{\ee}{\end{equation}}
\newcommand{\ba}{\begin{eqnarray}}
\newcommand{\ea}{\end{eqnarray}}

\newcommand{\Ms}{{\ensuremath{\mathrm{M}_{\odot}}}}

\newcommand{\Zs}{{\ensuremath{\mathrm{Z}_{\odot}}}}
\newcommand{\Ls}{{\ensuremath{\mathrm{L}_{\odot}}}}

\title{Abundant Water from Early Supernovae at Cosmic Dawn}

\author{Whalen, D. J.$^{*1}$, Latif, M. A.$^2$ and Jessop, C.$^1$}

\begin{document}

\maketitle

\begin{affiliations}

\item Institute of Cosmology and Gravitation, Portsmouth University, Dennis Sciama Building, Portsmouth PO1 3FX, \texttt{dwhalen1999@gmail.com}
\item Physics Department, College of Science, United Arab Emirates University, PO Box 15551, Al-Ain, UAE, \texttt{latifne@gmail.com}

\end{affiliations}

\begin{abstract} 

Primordial (or Pop III) supernovae were the first nucleosynthetic engines in the Universe, forging the heavy elements required for the later formation of planets and life. Water, in particular, is thought to be crucial to the cosmic origins of life as we understand it, and recent models have shown that water can form in low-metallicity gas like that present at high redshifts. Here we present numerical simulations that show that the first water in the Universe formed in Pop III core-collapse and pair-instability supernovae at redshifts $z \sim$ 20. The primary sites of water production in these remnants are dense molecular cloud cores, which in some cases were enriched with primordial water to mass fractions that were only a factor of a few below those in the Solar System today. These dense, dusty cores are also likely candidates for protoplanetary disk formation. Besides revealing that a primary ingredient for life was already in place in the Universe 100 - 200 Myr after the Big Bang, our simulations show that water was likely a key constituent of the first galaxies.

\end{abstract}

\begin{figure*}
\begin{center}
\includegraphics[width=\textwidth]{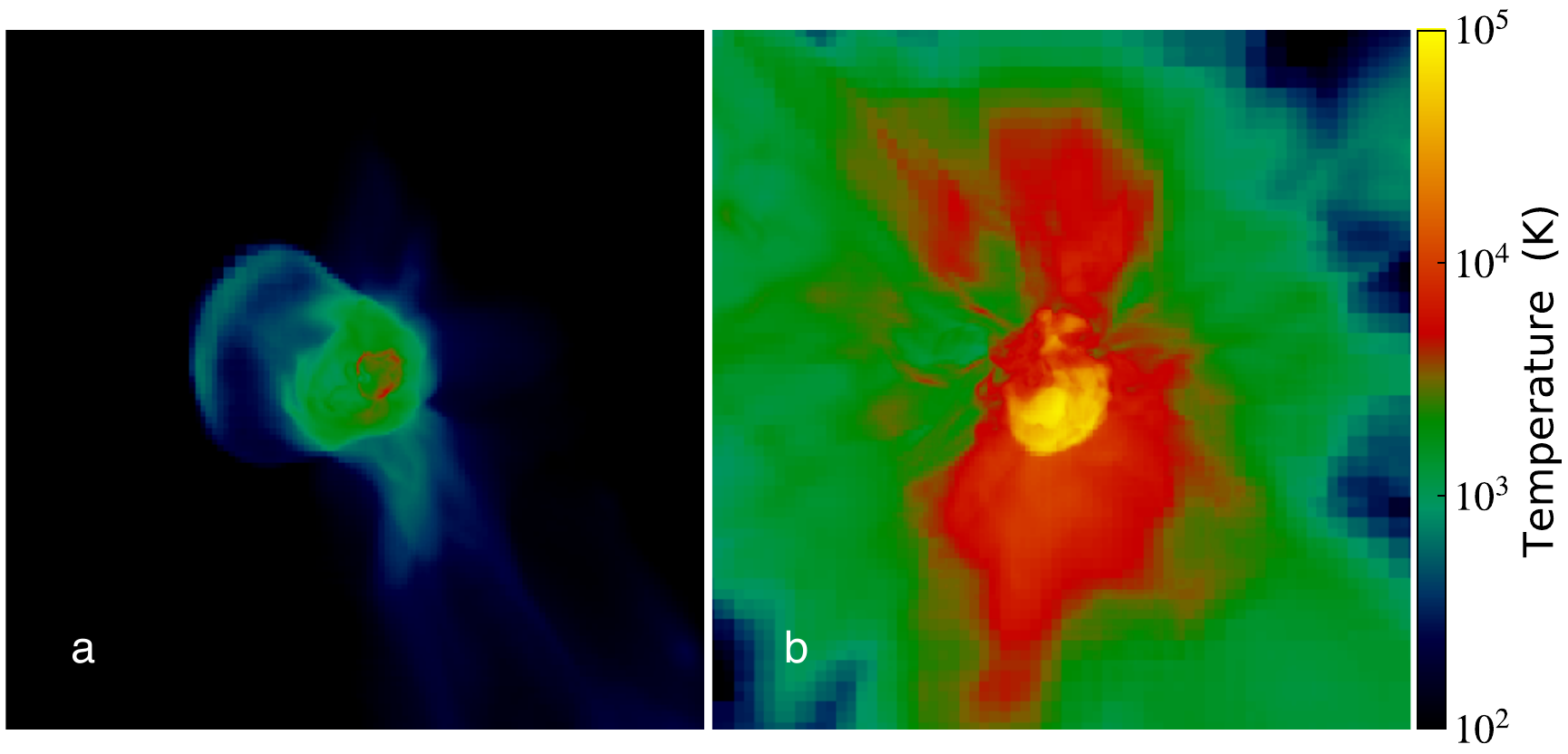} 
\end{center}
\caption{{\bf Primordial SN explosions.}  One kpc images of the 13 \Ms\ CC SN in the $1.1 \times 10^6$ \Ms\ halo 1.2 Myr after the explosion (a) and the 200 \Ms\ PI SN in the $2.2 \times 10^7$ \Ms\ halo 0.7 Myr after the explosion (b).  The relic H II regions of the stars are visible as the 2000 K - 10,000 K gas and the CC and PI ejecta are visible as the 10$^4$ and 10$^5$ K shocked gas with radii of $\sim$ 50 and 100 pc, respectively.  At the end of the simulations both SN remnants are still trapped within their respective H II regions.}
\label{fig:SNe}
\end{figure*}

We modeled the explosions of 13 \Ms\ (\Ms, solar mass) and 200 \Ms\ Pop III stars\cite{jet09b,hw10,wet12a} with the Enzo adaptive mesh refinement (AMR) code\cite{enzo}.  The 13 \Ms\ star forms when a cosmological halo grows to $1.1 \times 10^6$ \Ms\ at $z =$ 22.2.  It lives for 12.2 Myr and then explodes as a core-collapse supernova (CC SN) with an energy of 10$^{51}$ erg, ejecting 0.784 \Ms\ of metals with 0.051 \Ms\ of oxygen.  In the second simulation, the 200 \Ms\ star forms in a $2.2 \times 10^7$ \Ms\ halo at $z =$ 17.8. It lives for 2.6 Myr and then explodes as a pair-instability (PI) SN with an energy of $2.8 \times 10^{52}$ erg, ejecting 113 \Ms\ of metals with 55 \Ms\ of oxygen\cite{rs67,brk67,yosh22,xing23}.  Ionizing UV flux from both stars creates anisotropic H II regions with final radii of about 150 pc and 500 pc, respectively\cite{wan04,ket04,awb07}.  Neither ionization front breaks out of its halo so the explosions occur in trapped H II regions\cite{ritt12} with somewhat higher internal densities of $\sim$ 1 cm$^{-3}$.  After the stars explode, H$_2$, a key ingredient in water formation\cite{ssl15}, rapidly forms throughout their H II regions because it cools faster than it recombines.  As the SN shock sweeps up gas in the halo it cools, first by bremsstrahlung emission and then by collisional excitation and ionization of H and He\cite{ky05,wet08a}.  The two SNe are shown in Figure~\ref{fig:SNe}, where the relic H II regions are visible as the 2000 - 10,000 K gas and the CC and PI ejecta are visible as the 10$^4$ and 10$^5$ K shocked gas with radii of $\sim$ 50 and 100 pc, respectively. At the end of the simulations both SN remnants are still trapped within their H II regions.

\section{Water Synthesis}

\begin{figure*}
\begin{center}
\includegraphics[width=\textwidth]{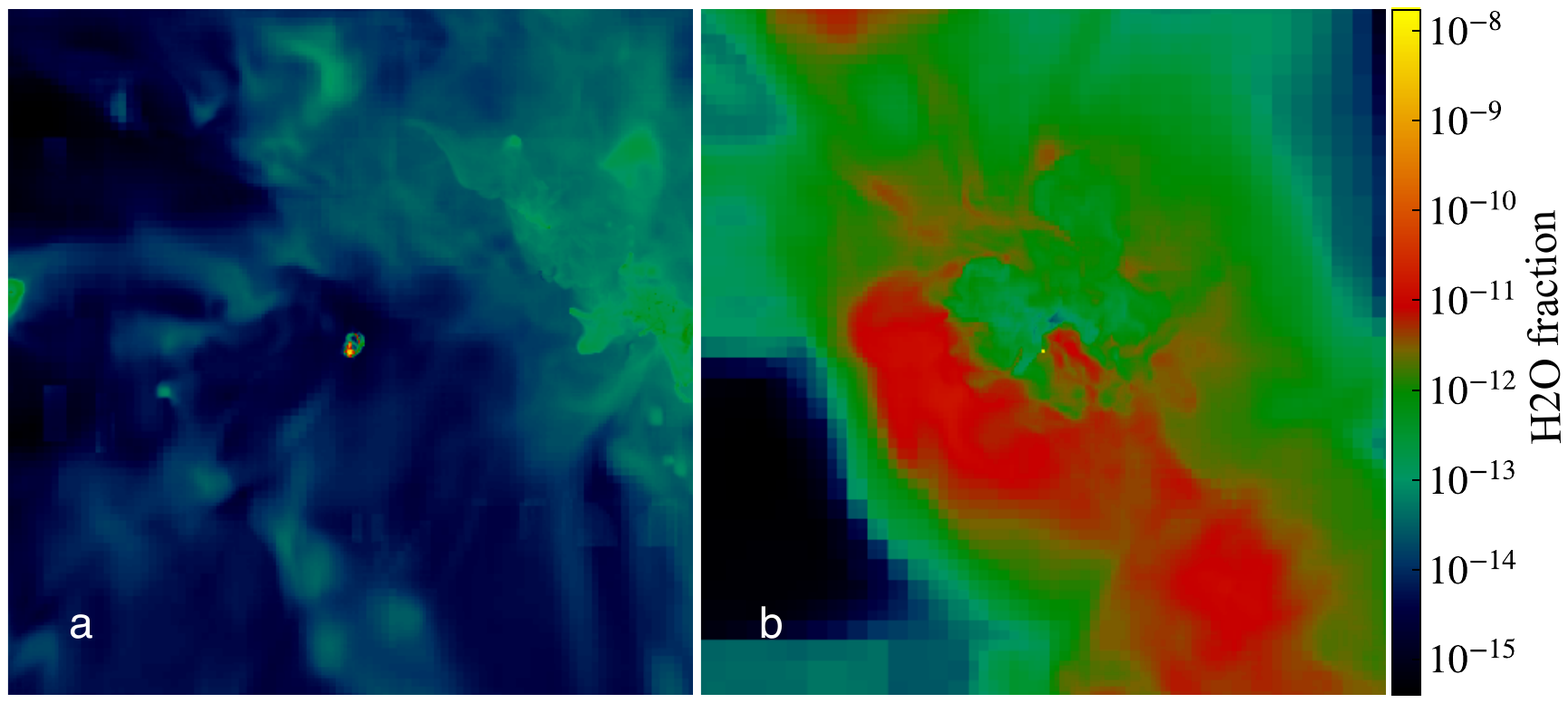} 
\end{center}
\caption{{\bf Water vapor in primordial halos.}   1 kpc images of water vapor in the 13 \Ms\ CC SN at 90 Myr after the explosion (a) and the 200 \Ms\ PI SN at 3 Myr after the explosion (b). Mass fractions for diffuse water vapor in the halos vary from 10$^{-14}$ - 10$^{-12}$ in the CC SN and 10$^{-12}$ - 10$^{-10}$ in the PI SN.  Dense clumps with much higher water masses are visible as the yellow specks in the centers of both images.}
\label{fig:WV}
\end{figure*}

\begin{figure}
\centering
\includegraphics[scale=0.7]{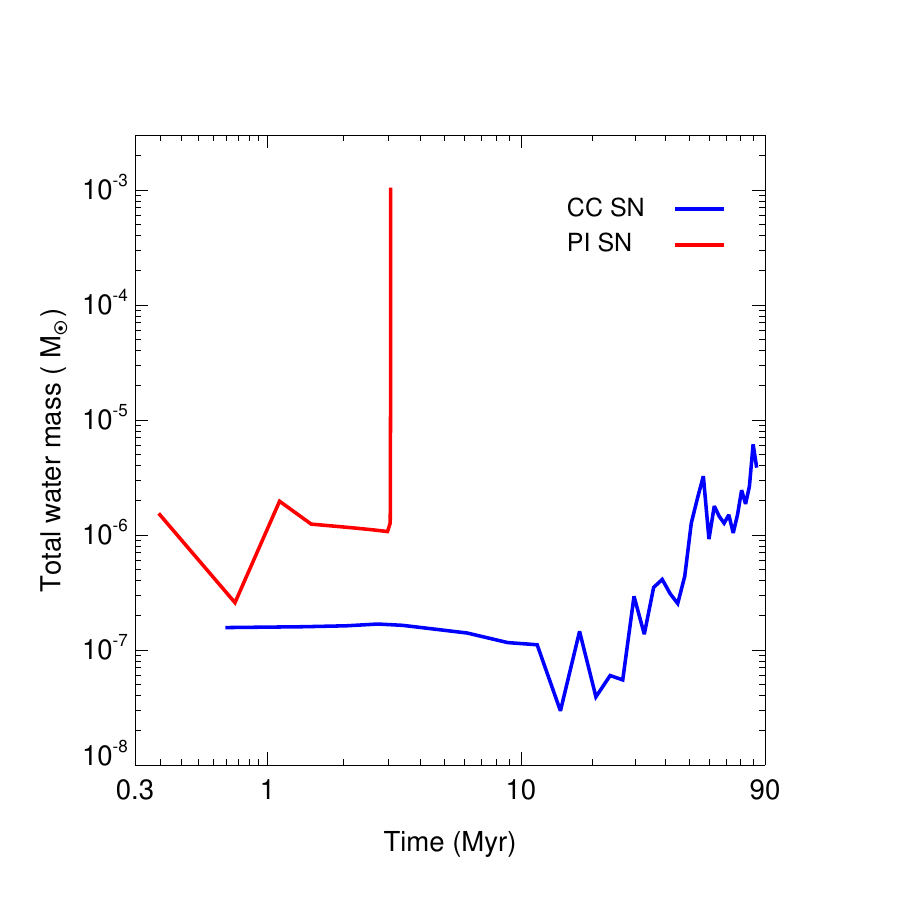} 
\caption{{\bf SN water masses.}   Total water masses in the CC (blue) and PI (red) SNe as a function of time since explosion, which are dominated by synthesis in dense cloud cores in their respective halos at late times.  Water formation rises sharply at earlier times in the PI SN core because cooling and collapse timescales are shorter at its higher metallicities.}
\label{fig:WM}
\end{figure}

As the SNe expand and cool, oxygen from the ejecta reacts with H and H$_2$ to form water in the halo, and H$_2$ also forms on dust grains.  As shown in Figure~\ref{fig:WV}, diffuse water vapor later permeates both halos with mass fractions of 10$^{-14}$ - 10$^{-12}$ in the CC SN and 10$^{-12}$ - 10$^{-10}$ in the PI SN.  Although water forms throughout both halos its total masses remain small and grow slowly over most of the simulation time, as seen in Figure~\ref{fig:WM}.  The small masses and slow growth are due to the relatively low densities in the expanding SN remnants, in which the reactions that produce water have low rates.  The water mass grows from 10$^{-8}$ - 10$^{-7}$ \Ms\ in the CC SN over the first 20 Myr and from 1 - 1.5 $\times$ 10$^{-6}$ \Ms\ in the PI SN over the first 2 - 3 Myr.  As shown in Figure~\ref{fig:WV}, water mass fractions in the PI SN on large scales are highest in the dense shell of gas that is swept up and chemically enriched by the expanding shock because densities, and thus H$_2$O reaction rates, are greatest there.

\begin{figure}
\centering
\begin{tabular}{cc}
\includegraphics[angle=90,width=0.5\textwidth]{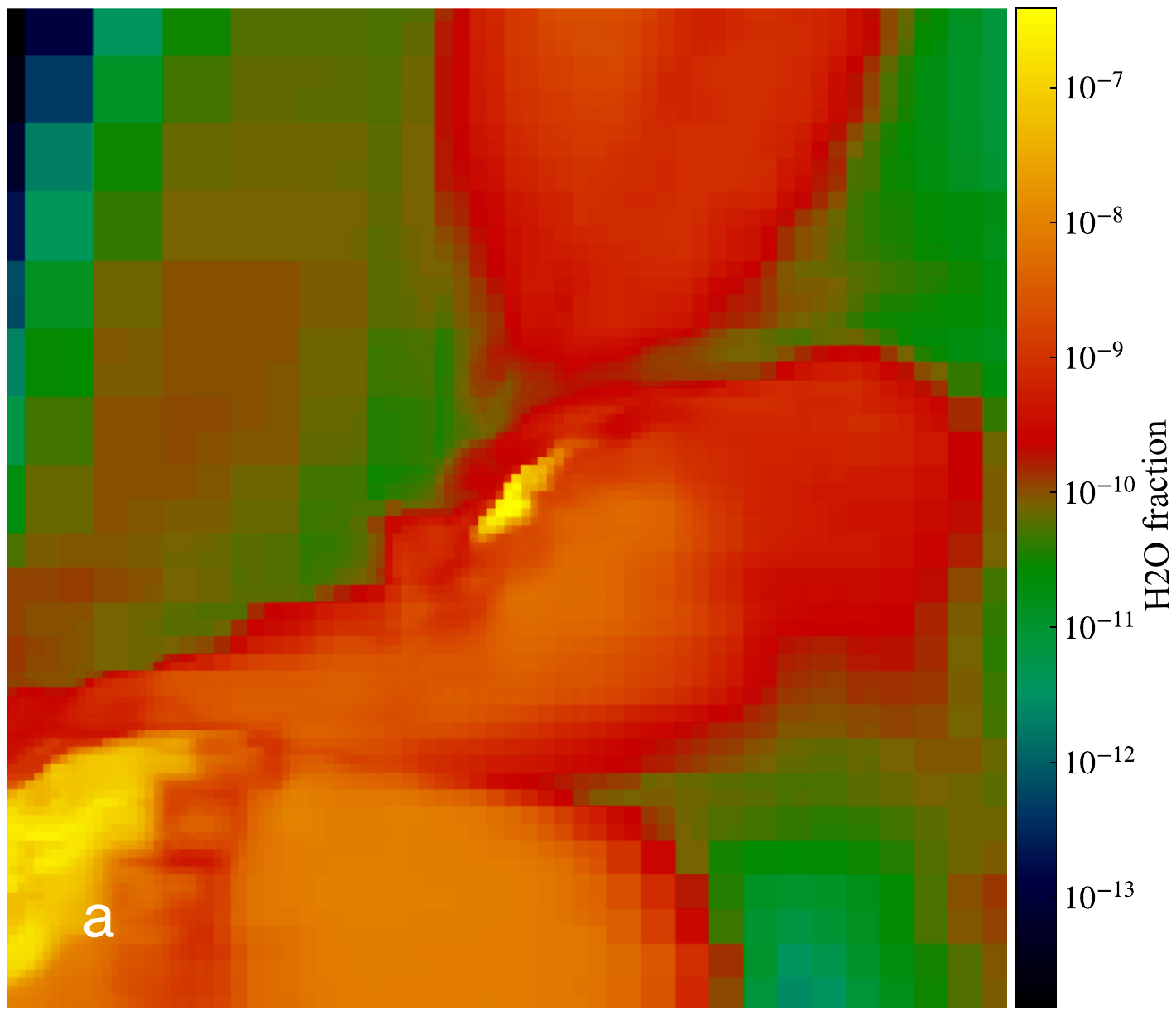} &
\includegraphics[angle=90,width=0.5\textwidth]{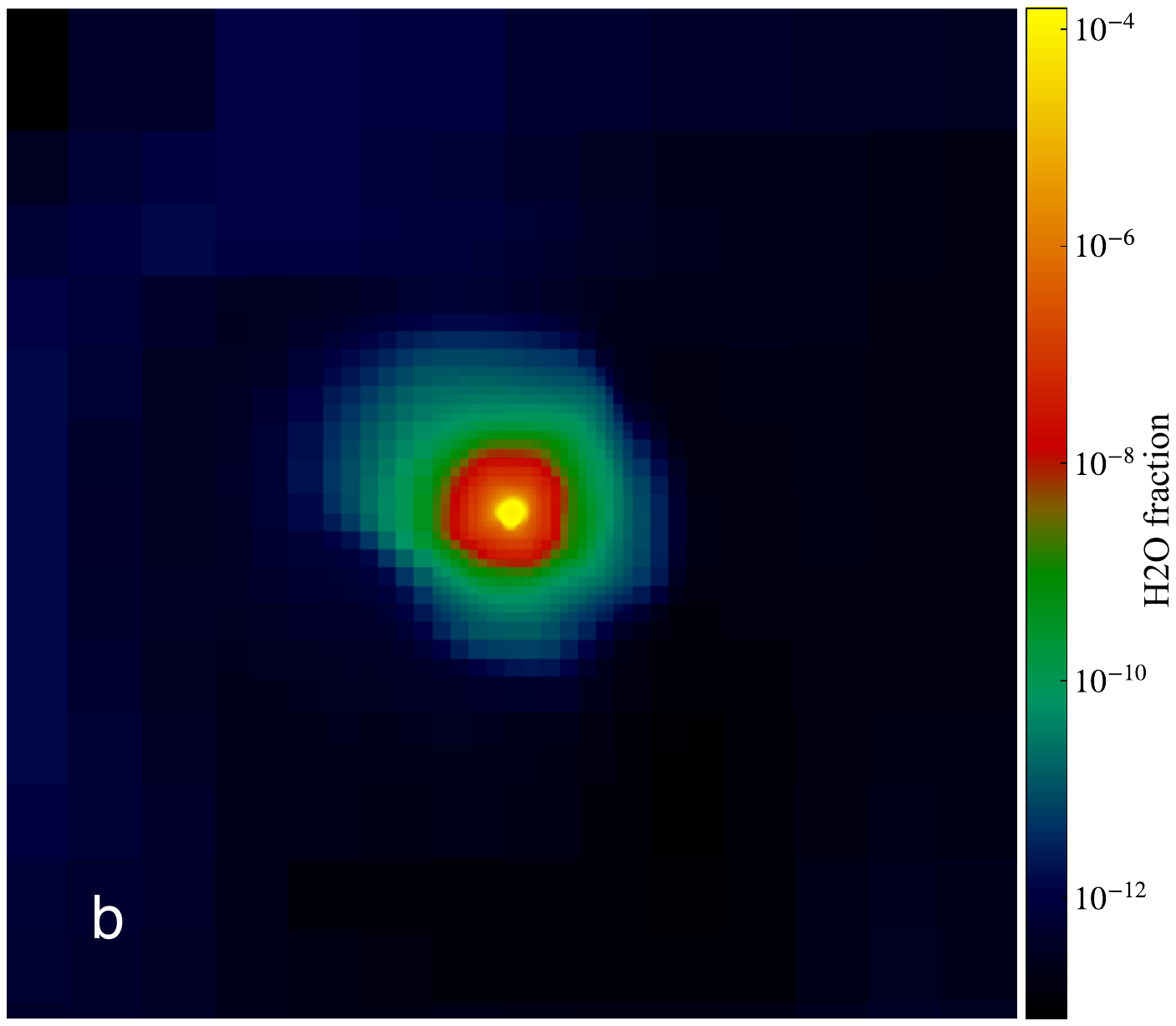} 
\end{tabular}
\caption{{\bf Water mass fractions in the dense cloud cores.}  3 pc and 0.1 pc images of water mass fractions in the CC SN core at 90 Myr (a) and the PI SN core at 3 Myr (b).}
\label{fig:WMF}
\end{figure}

But water masses then rise sharply by a few orders of magnitude in both halos, from 10$^{-6}$ \Ms\ - 10$^{-3}$ \Ms\ at 3 Myr in the PI SNR and from 10$^{-8}$ \Ms\ - 10$^{-6}$ \Ms\ from 30 - 90 Myr in the CC SNR.  This water forms almost entirely in two dense cloud cores, one in each halo, that were contaminated by metals from the explosions and then collapsed to high densities at which H$_2$O reaction rates abruptly rise.  As shown in Figure~\ref{fig:WMF}, water mass fractions reach 10$^{-4}$ in the PI SN fragment and $4 \times10^{-7}$ in the CC SN core by the end of the runs, the latter of which is consistent with those in one-zone models at similar metallicities and densities\cite{ssl15}.  The dominant sites of water production in primordial SNe are thus dense, self-gravitating cores in the ejecta\cite{slud16}, not the large volumes of diffuse gas enriched by water in the halo.

The CC SN core formed before the explosion and is gradually enriched by it.  As shown in Extended Data Figure~\ref{fig:merger}, turbulence in the wake of the merger of two halos at $z =$ 26.4 that later host the 13 \Ms\ star produces several gas clumps in its vicinity before its birth.  One was only 30 pc away, and after surviving photoevaporation by the star it collides with ejecta from the explosion 20 Myr later.  As shown in panel (a) in Extended Data Figure~\ref{fig:mix}, turbulence in the clump prior to the collision was transonic, with Mach numbers of $\sim$ 2, as expected for self-gravitating cores in the initial stages of collapse.  However, SN flows then buffet the clump, driving highly supersonic turbulence that gradually mixes it to metallicities $Z \sim$ 10$^{-4}$ \Zs\ (\Zs, solar metallicity) as shown in panel (b) of Extended Data Figure~\ref{fig:mix}.  It collapses to a radius of $\sim$ 0.1 pc at a mass of 1627 \Ms, a central density of $2.4 \times 10^8$ cm$^{-3}$, and a total water mass of $10^{-5}$ \Ms\ 90 Myr after the explosion.  

The PI SN core is created by the explosion.  As shown in the phase diagrams in Extended Data Figure~\ref{fig:phase}, the hot PI SN bubble\cite{mag20,mag22} promptly enriches surrounding gas to high (and even supersolar) metallicities at early times, as in previous cosmological simulations\cite{latif20c}.  Hydrodynamical instabilities in the expanding bubble produce turbulent density fluctuations that form a compact clump of gas at $Z =$ 0.04 \Zs.  It collapses to a radius of $\sim$ 0.01 pc at a mass of 35 \Ms, a central density of $6.0 \times 10^{14}$ cm$^{-3}$, and a total water mass of $9 \times 10^{-3}$ \Ms\ 3 Myr after the explosion.  At these densities dust cooling becomes important in the core, as shown in the phase plots in Extended Data Figure~\ref{fig:dust}. This clump becomes self-gravitating at much earlier times than in the CC SN because its higher metallicity results in faster cooling and collapse.  In contrast, the CC SN cloud core collapses on much longer timescales because of its much lower metallicities, and only after first being mixed with external metals by supersonic turbulence.  The CC SN core forms and collapses on timescales similar to those on which metals have formed second-generation stars in previous cosmological simulations\cite{ritt12,brit15,rit16,cw19}.  The PI SN core will form such stars far earlier than in any simulation to date.

\section{Discussion and Conclusion}

Cloud cores enriched by metals from Pop III SNe were probably the main sites of water formation in most primeval halos because similarities in explosion dynamics would have produced dense clumps across a wide range of energies, progenitor masses and halo masses.  Conditions that favored the formation of such cores, such as major mergers or explosions in trapped H II regions, thus maximized the production of primordial water.  SNe in compact H II regions tend to form clumps because dynamical instabilities in the expanding ejecta form sooner and the shock stalls at earlier times in higher ambient densities.  However, explosions can still form dense cores even if ionizing UV flux from the star breaks out of the halo and most of the baryons are lost to champagne flows because anisotropies in the H II region can still trap radiation along some lines of sight.

We considered just a single star forming in each halo as the simplest case.  Multiple stars may also form\cite{get11,sug20,prs21,latif22a,chon24}.  If so, several SN explosions may occur and overlap in the halo.  They may temporarily destroy water in low density regions but we expect the dense cores where most water forms to survive ionizing UV and SNe from other stars, just as the dense clump 30 pc from the 13 \Ms\ star survived its radiation and explosion.  Multiple explosions may produce more dense cores and, thus, more sites for water formation and concentration in the halo.  

The highest redshift at which water has been found to date is $z =$ 6.9 by detections of p-H$_2$O(2$_{1,1}$ - 2$_{0,2}$) and p-H$_2$O(3$_{1,2}$ - 2$_{1,2}$) transition lines at 752 GHz and 1153 GHz with the Atacama Large Millimeter Array (ALMA)\cite{jaru21}. H$_2$O(4$_{2,3}$ - 4$_{1,4}$) and H$_2$O(3$_{3,0}$ - 3$_{2,1}$) lines at 2264 GHz and 2196 GHz have also been discovered by ALMA in SPT 0346-52 at $z =$ 5.656\cite{jones19}.  These lines, which are pumped by far infrared (FIR) emission from dust, would still be redshifted into ALMA bands at $z \gtrsim$ 15.  Gas and dust temperatures in our models exceed those that activate these transitions ($\sim$ 400 K) so primordial halos emitted these lines.  When the cores later form stars, dust can reprocess their radiation and populate metastable levels in H$_2$O that produce maser emission at 22 GHz.  With luminosities of 10$^{-3}$ - 10$^3$ \Ls\ (\Ls, solar luminosity), redshifted line emission from individual masers at $z \sim$ 20 would probably not be visible to the Square Kilometer Array or Next Generation Very Large Array today.  But a global population of these masers might have produced a cosmic line background at the end of the cosmic Dark Ages that could be found by these observatories in the coming decade.  The same is true of FIR-pumped millimeter lines from halos with ALMA. 

Recent numerical simulations of exoplanet formation down to the lowest metallicities ever attempted indicate that both cores are probable sites of planet formation.  Gas in the $Z =$ 10$^{-4}$ \Zs\ CC SN clump could produce protoplanetary disks that fragment into Jupiter-mass planets\cite{met22}.  The higher metal content of the $Z =$ 0.04 \Zs\ PI SN fragment could in principle lead to the formation of rocky planetesimals in protoplanetary disks with low-mass stars.  This latter point is corroborated by the fact that Jeans masses fall to 1 - 2 \Ms\ at the center of the PI SN core, as shown in Extended Data Figure~\ref{fig:jmass}.  We would not expect such planets  to have much impact on exoplanet demographics today (except to extend its low-metallicity tail to smaller values) because Pop III stars were relatively sparse at $z \sim$ 15 - 20, but they could be detected as extinct worlds around ancient, metal-poor stars in the Galaxy in future exoplanet surveys\cite{bvd24}.  

Our Enzo simulations show that these disks would have been heavily enriched by primordial water, to mass fractions that were 10 - 30 times greater than those in diffuse clouds in the Milky Way in the CC SN core and to only a factor of a few lower than those in the Solar System today in the PI SN core.  The large H$_2$O mass fractions and the potential for low-mass star formation in the PI SN core raise the possibility of a habitable zone in the protoplanetary disk in which equilibrium temperatures allow water to exist in liquid form\cite{1993Kasting,2023ASPC..534.1031K,2014A&A...567A.133V}.  If planetesimals can form at $Z =$ 0.04 \Zs\ in the disk, the planets into which they grow could harbor water.   

Our simulations suggest that water was present in primordial galaxies because of its prior formation in their constituent halos\cite{jeon11,wise12,ren15,latif22b}.  Water mass fractions in diffuse SN remnants taken up into these galaxies could reach 10$^{-10}$, only an order of magnitude less than in the Milky Way today.  Some of this water would have been photodissociated by massive, low-metallicity stars in these galaxies (see Extended Data) or destroyed by additional chemical reactions as they reached higher metallicities at later times.  But rising dust fractions in early galaxies would also have shielded water from UV and mitigated its destruction to some degree.  How much water survived the harsh radiation environments of the first galaxies remains to be determined.

%\bibliography{refs}

% Fig. 1

% Fig. 2

% Fig. 3

% Fig. 4

\newpage

\noindent {\bf Methods}    

The Enzo cosmology code\cite{enzo} has an N-body particle-mesh scheme\cite{efs85,couch91} for evolving dark matter that is self-consistently coupled to hydrodynamics, nonequilibrium primordial gas chemistry, and ionizing UV transport with the MORAY raytracing radiation code\cite{moray}.  Our simulations use the piecewise parabolic method for hydrodynamics\cite{wc84,bryan95} and the HLLC scheme for enhanced stability with strong shocks and rarefaction waves\cite{toro94}.  We use four energy bins in MORAY:  H and He ionizing photons, H$^-$ photodetachment photons and Lyman-Werner (LW) photons. 

We implemented nonequilibrium water chemistry\cite{cw19} in Grackle\cite{grackle}, with 49 reactions in 15 primordial gas species (e$^-$, H, H$^+$, H$_2$, H$^-$, H$_2^+$, HeH$^+$, He, He$^+$, He$^{2+}$, D, D$^+$, D$^-$, HD, HD$^+$, HeH$^+$, D$^-$ and HD$^+$) and 40 reactions in 17 metal and molecular species (C$^+$, C, CH, CH$_2$, CO$^+$, CO, CO$_2$, O$^+$,O, OH$^+$, OH, H$_2$O$^+$, H$_2$O, H$_3$O$^+$, O$_2^+$, O$_2$, Si, SiO, and SiO$_2$)\cite{c18}.  The 40 reactions are Z1 - Z35 and Z40 from Table 1 of Omukai et al (2005)\cite{om05} and 
\begin{center}
\begin{tabular}{ccc}
O$^+$  $+$   e$^-$    &  $\rightarrow$  &       O         $+$  $\gamma$   \\
Si         $+$  O$_2$   &  $\rightarrow$  &     SiO       $+$          O         \\
Si         $+$    OH      &  $\rightarrow$  &      SiO       $+$         H          \\   
SiO      $+$    OH      &  $\rightarrow$  &  SiO$_2$   $+$         H          \\   
\end{tabular}
\end{center}
from Table 1 of Chiaki et al (2015)\cite{c15}.

We evolve the reaction network with a fully implict scheme.  It includes updates to the gas energy due to heat from H$_2$ formation and collisional excitation and ionization cooling by H and He, recombination cooling, bremsstrahlung cooling, and inverse Compton cooling by the CMB at $T >$ 8000 K. We also include H$_2$ and HD line cooling for $T <$ 10,000 K, cooling due to fine structure emission by C$^+$, C and O\cite{ss06}, and cooling by transitions between rotational levels in OH, H$_2$O and CO, where the line rates are obtained from interpolations in precomputed tables\cite{nk93,nlm95,ohy10}.  Corrections to optically thin gas cooling by metal and molecule lines at high densities are also included\cite{cw19}.  We note that HD formation is evident in the relic H II region of the PI SN remnant in Extended Data Figures~\ref{fig:dust} and \ref{fig:water}, where it cools gas down to the CMB temperature at that redshift.

UV photons can create and destroy H$_2$O and its precursors, e.g.,\cite{bs15} 
\begin{center}
\begin{tabular}{cccc}
H$_2$O  $+$  $\gamma$  &  $\rightarrow$  &  H$_2$  $+$  O    &  (h$\nu$ $>$ 9.5 eV)   \\
H$_2$O  $+$  $\gamma$  &  $\rightarrow$  &  H          $+$  OH  &  (h$\nu$ $>$ 6.0 eV)   \\
OH          $+$  $\gamma$  &  $\rightarrow$  &  O          $+$  H    &  (h$\nu$ $>$ 6.4 eV).   
\end{tabular}
\end{center}
However, there are no other stars in the halo or its vicinity that could photodissociate water so we do not include these reactions in our simulations.  In principle there could be diffuse UV emission from the remnant itself but these fluxes would be much lower than those of nearby stars.  While these reactions would be required if additional stars formed in the halo there is enough dust in the cores to mitigate their effects to some degree.  We do include photodissociation of H$_2$ by 11.18 eV - 13.6 eV LW UV photons and photodetachment of H$^-$ by continuum photons above 0.755 eV.  

H$_3$O$^+$ formation by cosmic rays (CRs) is an important pathway to water formation in the Galaxy today.  However, we expect CR densities in the primordial Universe to be much lower than in the Milky Way because Pop III star formation was relatively sparse.  Consequently SNe, which produce CRs via first- and second-order Fermi acceleration in the shock, had low event rates so $z \sim$ 20 was too early for a strong CR background to have arisen. We therefore exclude the H$_2$O formation via the H$_3$O$^+$ channel in our simualtions.  However, the PI and CC SNe themselves produced some CRs so we somewhat underestimate H$_2$O mass fractions in the diffuse gas and dense cores. 

Dust formation and destruction and gas cooling due to thermal emission from dust grains\cite{cw19} is also included in our simulations\cite{n03,n07,c15}.  It includes eight dust species:  metallic silicon (Si), metallic iron (Fe), forsterite (Mg$_2$SiO$_4$), enstatite (MgSiO$_3$), amorphous carbon (C), silica (SiO$_2$), magnesia (MgO), and troilite (FeS).  Our model includes the ten chemical reactions that create these species from Table 2 of Chiaki (2015)\cite{c15} along with
\begin{center}
\begin{tabular}{ccc}
CO  $+$  2H$_2$    &  $\rightarrow$  &  CH$_3$OH             \\
H$_2$O$_{\rm (g)}$  &  $\rightarrow$  &  H$_2$O$_{\rm (s)}$,
\end{tabular}
\end{center}
where (g) and (s) refer to gas and solid phases, respectively.  Five of the 12 reactions deplete water onto dust grains while rapid catalysis of H$_2$ by dust grains promotes water formation\cite{om05}.  Dust grains do not directly catalyze water formation at $z \sim$ 20 because the cosmic microwave background maintains them at temperatures of 60 K, which are too high for reactants to bind to their surfaces.  

Our CC SN dust yields, compositions and grain size distributions are taken from Chiaki \& Wise (2019)\cite{cw19}, which are from Nozawa et al. (2007)\cite{n07}.  The CC SN grain size distribution varies as $r^{-3.5}$, similar to that in the Milky Way today\cite{pol94}. This distribution and our dust composition are shown in Figures 3 and 2 of Chiaki \& Wise (2019), respectively.  Our PI SN yields and grain sizes were taken from Nozawa et al. (2007).  They solved detailed nucleation models to obtain dust grain radii that included destruction due to thermal sputtering in the reverse shock of the SN remnant.  Because the strength of the reverse shock in part is determined by ambient densities, their dust yields are parametrized by H II region density.  We adopted grain yields for 13 \Ms\ and 200 \Ms\ CC and PI SN explosions in ambient densities of 1 cm$^{-3}$ from Nozawa et al. (2007), as these densities were closest to those in their respective H II regions in Enzo.  These dust mass fractions are relatively low because the nucleation models predict that up to 90\% of the dust originally formed in the explosion will be destroyed by the reverse shock at these local densities.  However, these and other calculations of grain growth in SNe\cite{bian07} are based on 1D Lagrangian explosion models that exclude hydrodynamical instabilities in 3D that can lead to the formation of dense clumps like the two studied here.  Hydrodynamical studies have since shown that clumping in actual reverse shocks in 3D can shield dust from thermal sputtering\cite{slav20}.  Consequently, our dust mass fractions should be taken as (possibly severe) lower limits.

We initialize our CC SN and PI SN simulations in 1 $h^{-1}$ Mpc and 1.5 $h^{-1}$ Mpc boxes at z = 200 with cosmological initial conditions generated with MUSIC\cite{hahn11} from the second-year Planck best fit lowP $+$ lensing $+$ BAO $+$ JLA $+$ H0 cosmological parameters:  $\Omega_{\mathrm M}$ = 0.3089, $\Omega_{\Lambda}$ = 0.691, $\Omega_{\mathrm b}$ = 0.0486, $\sigma_{\mathrm 8} =$ 0.816, $h^{-1} =$ 0.677, and $n =$ 0.967\cite{planck2}.  We first performed low-resolution 256$^3$ unigrid DM-only runs to select halos for the two stars.  In the CC SN run we then centered two nested grids on the halo that spanned 10\% of the top grid and evolved it with full baryonic physics down to 20 levels of refinement and 4 zones per Jeans length for a maximum resolution of 2063 AU.  In the PI SN run we centered three nested grids on the halo that spanned 10\% of the top grid and evolved it with full baryonic physics with up to 28 levels of refinement and 16 zones per Jeans length to achieve a maximum resolution of 2.1 AU.  In the PI SN run we included a LW UV background of 100 J$_{21}$, where J$_{21} =$ 10$^{-21}$ erg$^{-1}$ s$^{-1}$ Hz$^{-1}$ steradian$^{-1}$, that delayed star formation until the halo grew to a a little above 10$^7$ \Ms.

We chose 13 \Ms\ and 200 \Ms\ progenitors because they lie near the center of the mass ranges expected for these events, 8 - 20 \Ms\ for most CC SNe and 140 - 260 \Ms\ for PI SNe.  The energy of the CC SN, 10$^{51}$ erg, also lies near the middle of those observed for most of these events, 0.6 - $2.4 \times 10^{51}$ ergs while the PI SN energy is an emergent feature of thermonuclear burning of O and SI in the stellar evolution models that produce the explosion.  Elemental abundances for the CC SN are shown in Figure~1 of Chiaki \& Wise (2019)\cite{cw19}.  The PI SN abundances exhibit the usual 'odd-even' effect, in which even numbered nuclei are preferentially synthesized over odd numbered nuclei by 1 - 2 dex.  Energies and nucleosynthetic yields for both SNe are taken from precomputed databases\cite{hw02,hw10}. 

\subsection{Water formation pathways during collapse.}

The primary H$_2$O formation channels during collapse of the PI SN core are reactions Z10 - Z12,
\begin{center}
\begin{tabular}{cccc}
Z10:   &    O    $+$     H       &  $\rightarrow$  &  OH           $+$  $\gamma$      \\
Z11:   &    O    $+$  H$_2$  &  $\rightarrow$  &  H              $+$        OH            \\
Z12:   &  OH   $+$  H$_2$  &  $\rightarrow$  &  H$_2$O   $+$          H    .   
\end{tabular}
\end{center}
Z10 is more important at early stages of collapse but Z11 takes over OH formation at $n \sim$ 10$^8$ -10$^{10}$ cm$^{-3}$ when 3-body formation of H$_2$ begins to molecularize the core.  At $Z =$ 10$^-3$ \Zs\ in previous idealized one-zone models of collapse\cite{om05} most of the O goes into O$_2$ during collapse via reaction Z19,
\begin{center}
\begin{tabular}{cccc}
Z19:   &    O    $+$     OH       &  $\rightarrow$  &  O$_2$          $+$  H      
\end{tabular}
\end{center}
because temperatures in those models stay below below 300 K. But this is not so in our PI SN core, where compressional and shock heating keeps gas above 300 K (and closer to 1000 K most of the time), in spite of its high metallicity, 0.04 \Zs, as shown in Extended Data Figure~\ref{fig:water}.  There and in Extended Data Figure~\ref{fig:wmfrac} it can be seen that central water mass fractions level off at $\sim$ 10$^{-4}$ at densities above 10$^{10}$ cm$^{-3}$, after three-body production has mostly molecularized the core. We also show the evolution of the ratio of H$_2$O/O mass fractions versus central density during collapse in Extended Data Figure~\ref{fig:h2ofrac}, where it is seen that H$_2$O formation dominates O$_2$ formation above $n \sim$ 10$^{10}$ cm$^{-3}$.  Z10 - Z12 also dominate H$_2$O formation in the CC SN core.  While we can only follow its collapse to central densities $n \sim$ 10$^8$ cm$^{-3}$ most of the O goes into H$_2$O instead of O$_2$ because of the lower metallicity, $\sim$ 10$^{-4}$ \Zs, consistent with Figure 5c of Omukai et al. 2005.

\noindent {\bf Data Availability} \hspace{0.1in} The Enzo parameter files and initial conditions files generated by MUSIC that are required to perform the simulations are available at https://doi.org/10.5281/zenodo.5853118 (DOI: 10.5281/zenodo.5853118).  The MUSIC input files required to generate the initial conditions are available at https://sites.google.com/site/latifmaastro/ics.  

\noindent {\bf Code Availability} \hspace{0.1in} The code used to produce our cosmological simulations, Enzo 2.6, can be found at https://bitbucket.org/enzo/enzo-dev/tree/enzo-2.6.1.

\begin{addendum}

\item 

MAL was supported by UAEU UPAR grant No. 31S390.  CJ was supported by STFC grants ST/S505651/1 and ST/T506345/1.  The Enzo simulations and yt analyses were performed on HPC resources at UAEU and the Institute of Cosmology and Gravitation at the University of Portsmouth (Sciama).  

\item[Author Contributions] DJW proposed and developed this study, helped interpret its data, supervised CJ, and wrote the paper. MAL helped develop this study, supervised CJ, performed the Enzo PI SN simulations, and analyzed its data.  CJ helped perform the Enzo PI SN simulations.    

\item[Competing Interests] The authors declare no competing interests.  

\item[Supplementary videos]are available in the online version of this paper.
\item[Reprints and permissions information]is available at www.nature.com/reprints.
\item[Correspondence and requests]for materials should be addressed to DJW or MAL.   

\end{addendum}

\setcounter{figure}{0}

\renewcommand{\figurename}{Extended Data Figure}
\renewcommand{\tablename}{Table~Extended Data}

% EDF 1 

\begin{figure*}
\begin{center}
\begin{tabular}{cc}
\includegraphics[width=0.455\textwidth]{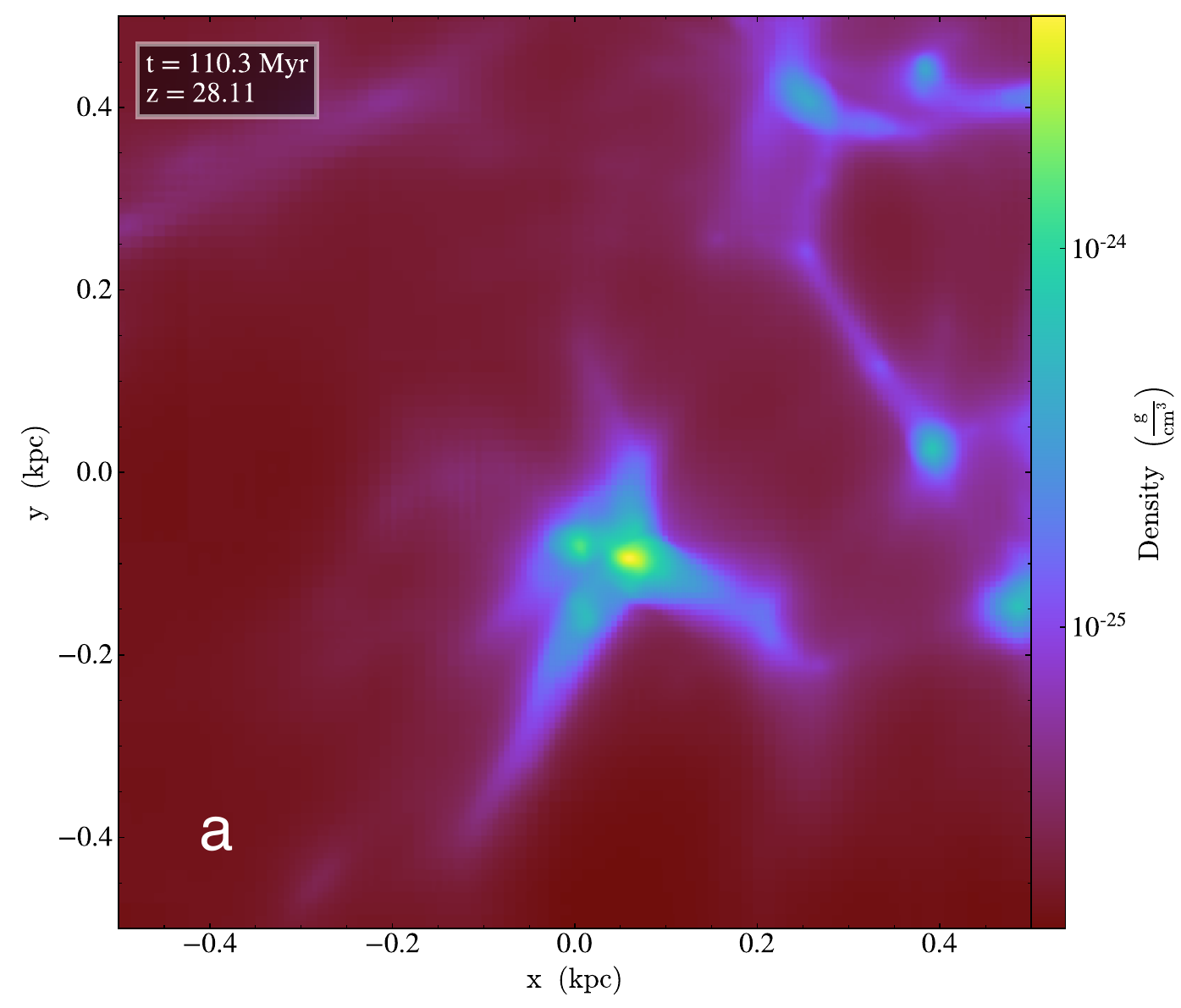} &
\includegraphics[width=0.5\textwidth]{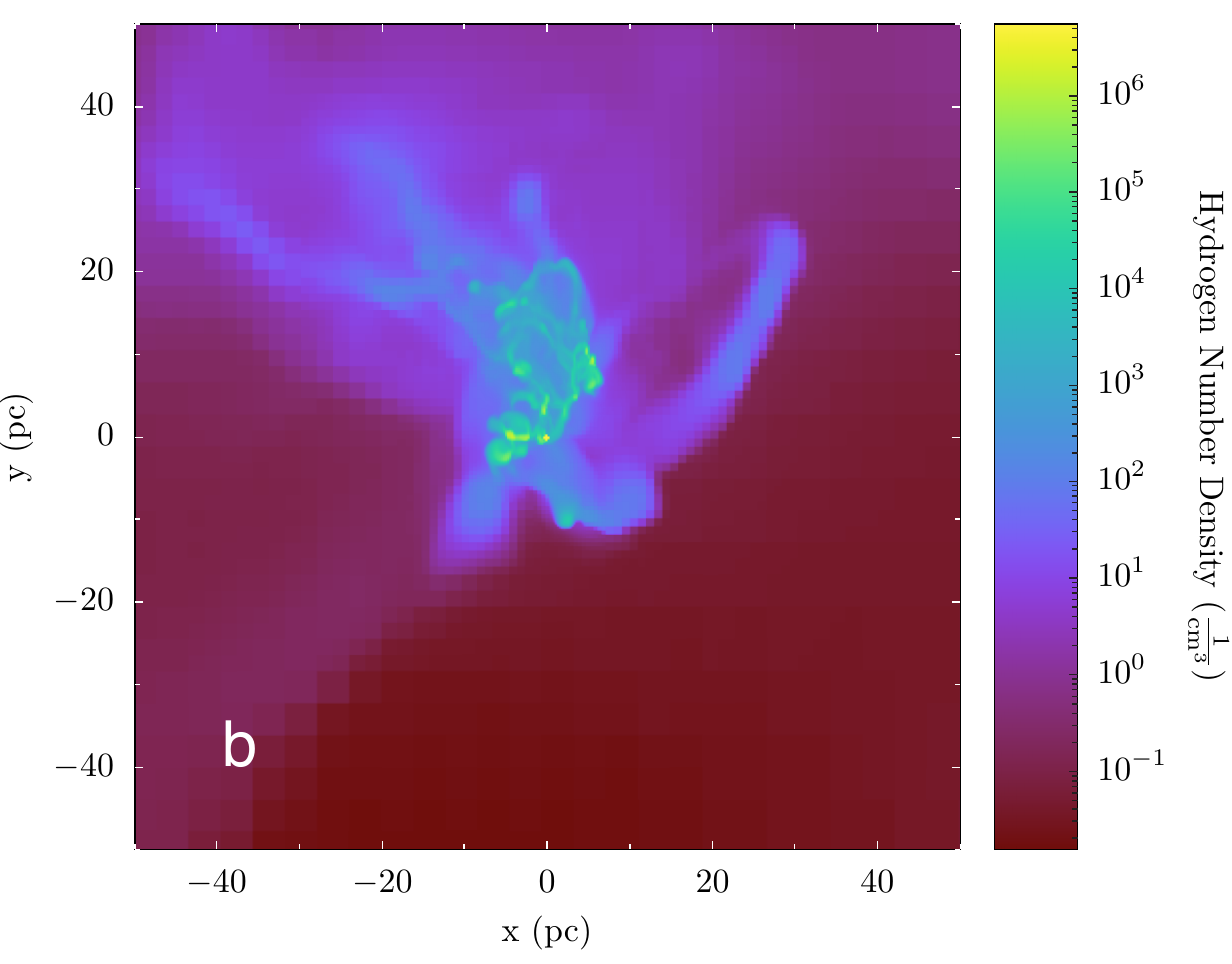} 
\end{tabular}
\end{center}
\caption{{\bf Clump formation in the CC SN halo.}  Left:  the $2 \times 10^5$ and $4.6 \times 10^4$ \Ms\ halos at $z =$ 28.11 (center) just before merging at $z =$ 26.4 and later growing to $1.1 \times 10^6$ \Ms\ and forming the 13 \Ms\ star at $z = $ 22.2. Right: zoom-in of the merged halo showing dense clumps created by turbulence during the encounter.}
\label{fig:merger}
\end{figure*}

% EDF 2 

\begin{figure}
\centering
\begin{tabular}{cc}
\includegraphics[width=0.485\textwidth]{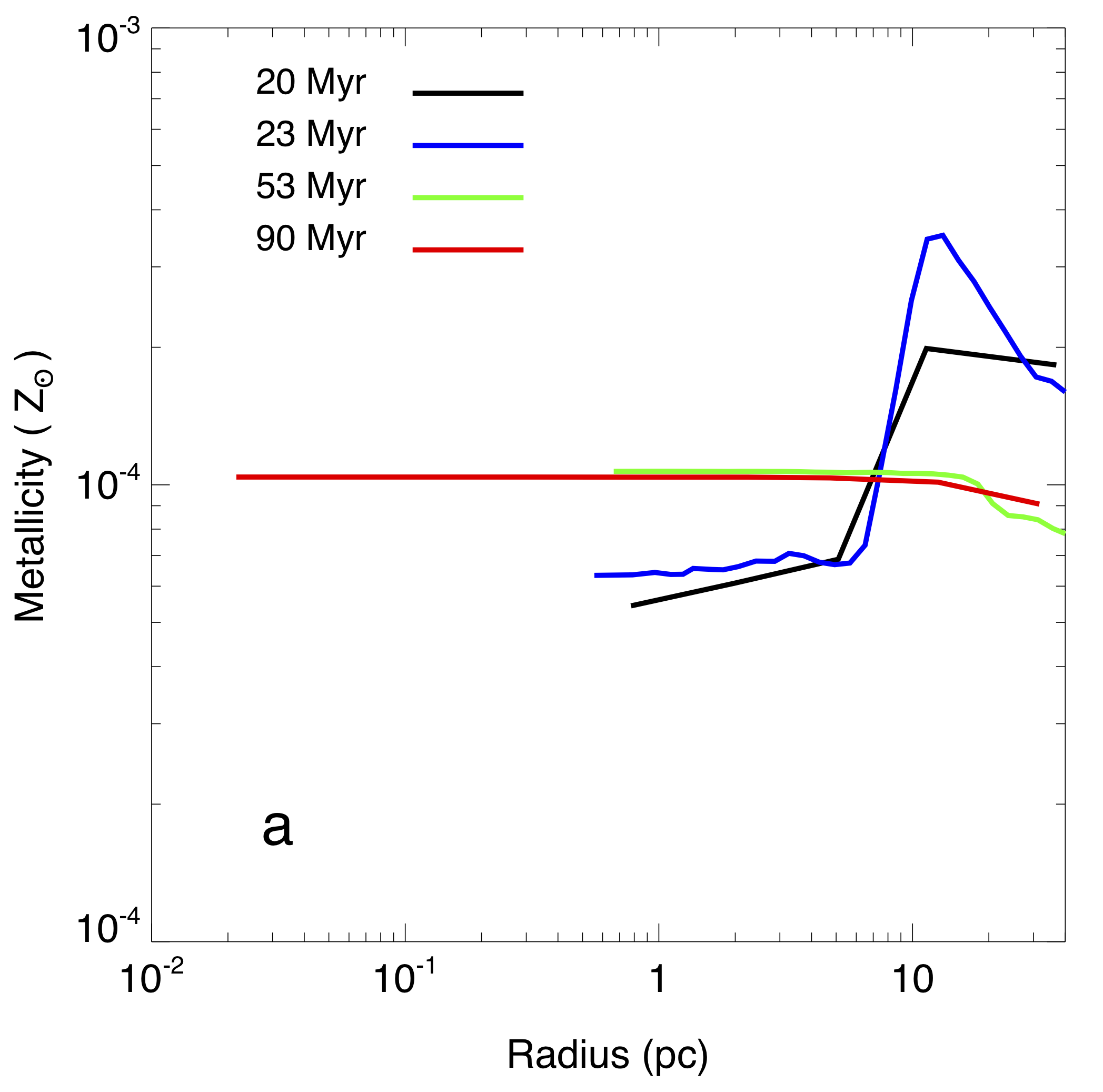} &
\includegraphics[width=0.5\textwidth]{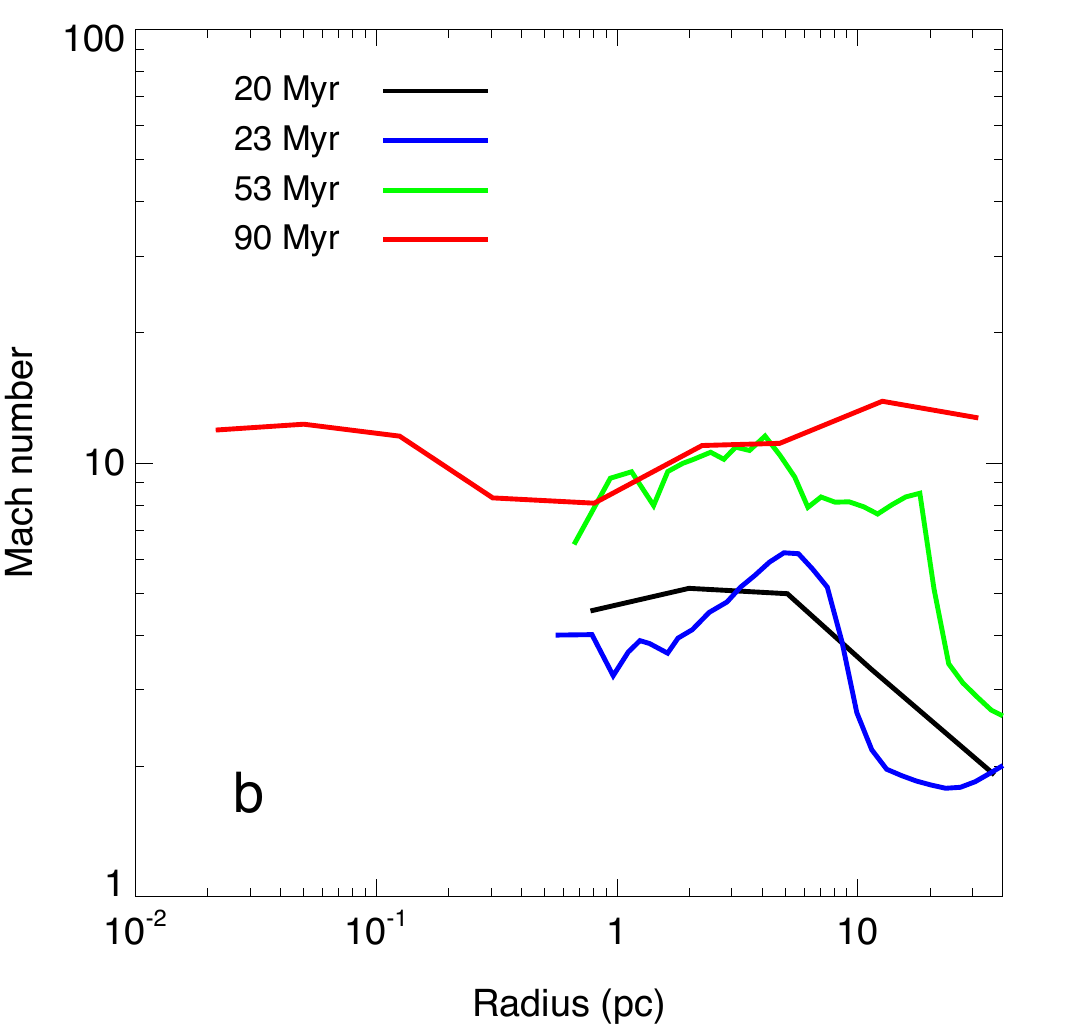} 
\end{tabular}
\caption{{\bf Enrichment of the clump by the CC SN.}  (a)  Infusion of the clump with metals from the explosion at 20 Myr, 23 Myr, 53 Myr and 90 Myr.  (b) Supersonic turbulence in the clump.  At 20 Myr, the SN shock has just crashed into the clump and metals have begun to pile up in its outer layers at radii of 5 - 10 pc, with some reaching depths of $\sim$ 0.7 pc.  Relic transonic turbulence from the initial stages of collapse prior to the collision is visible in the low Mach numbers at 20 - 30 pc. The abrupt conversion of the bulk kinetic energy of inflow into random motions upon collision with the clump drives the supersonic turbulence at 5 - 10 pc, where the Mach numbers reach 5.  At 23 Myr even more metals have piled up in the outer regions of the core and its internal metallicities have slightly risen.  From 53 - 90 Myr the flows cause the entire core to become highly supersonically turbulent, and the metals that had previously accumulated in its outer regions now fully permeate it.}
\label{fig:mix}
\end{figure}

% EDF 3 

\begin{figure}
\centering
\begin{tabular}{cc}
\includegraphics[angle=90, origin=c, width=0.5\textwidth]{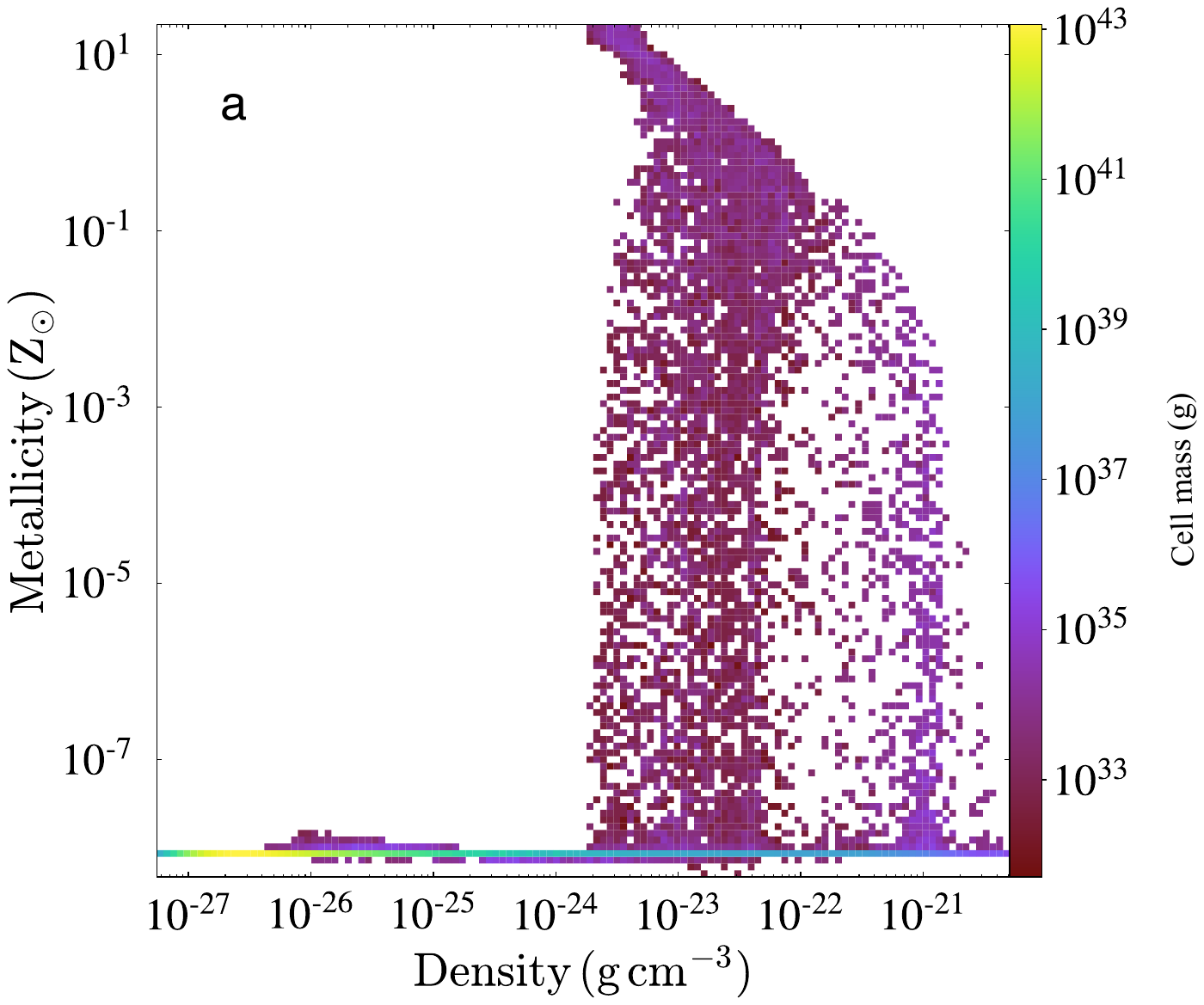} &
\includegraphics[angle=90, origin=c, width=0.5\textwidth]{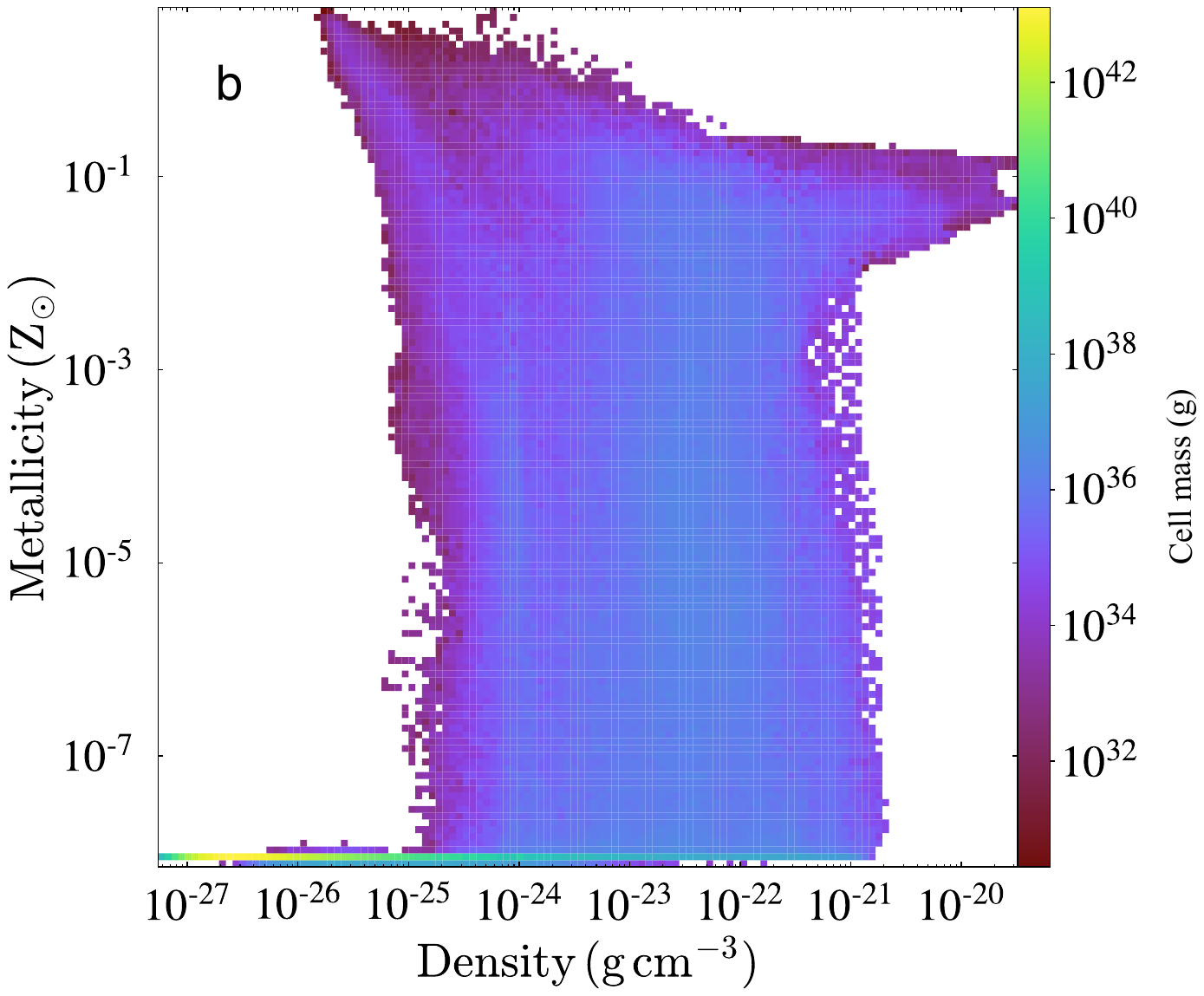} \\
\includegraphics[angle=90, origin=c, width=0.5\textwidth]{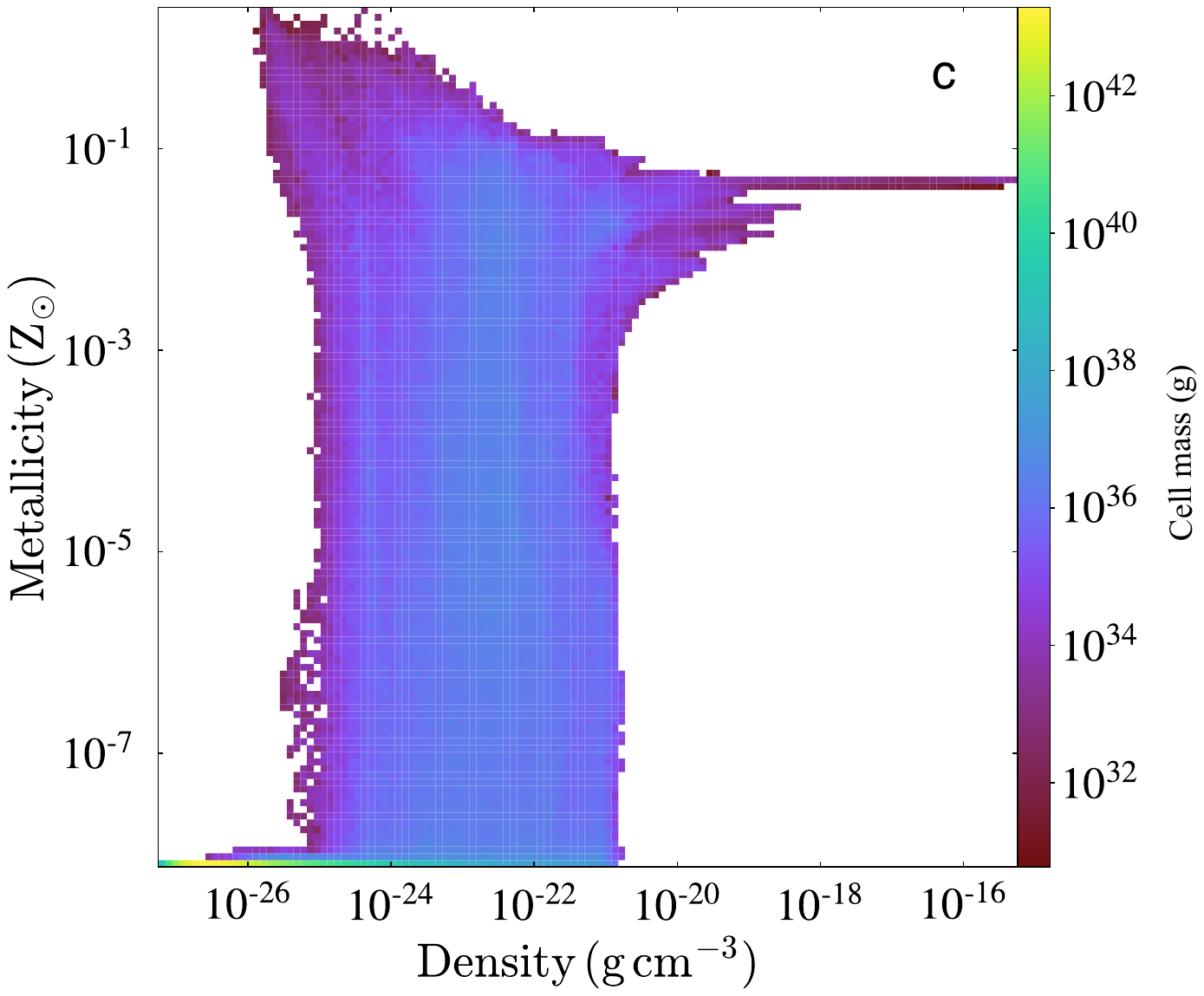} &
\includegraphics[angle=90, origin=c, width=0.5\textwidth]{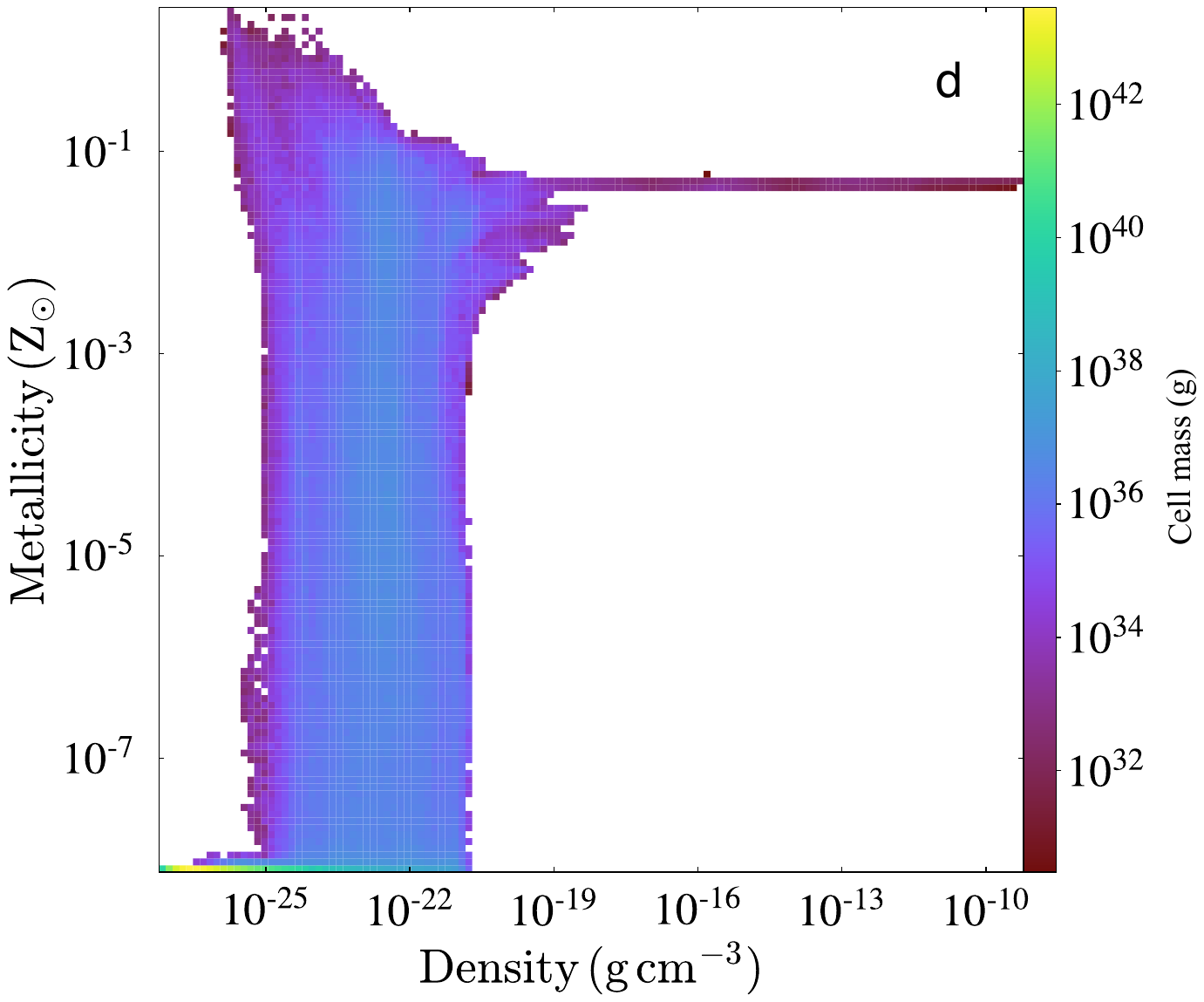} 
\end{tabular}
\caption{{\bf Emergence of the PI SN clump from $Z =$ 0.04 \Zs\ gas.}  (a) The initial rapid enrichment of gas in the PI SN bubble 0.333 Myr after the explosion, in which the dense shell swept up by the expanding remnant (at $\sim$ 10$^{-22}$ g cm$^{-3}$) has already reached metallicities $Z \sim$ 10$^{-2}$ - 10$^{-3}$ \Zs.  (b)  The appearance of a turbulent density fluctuation at $\sim$ 10$^{-21}$ g cm$^{-3}$ and $Z \sim$ 0.1 - 0.01 \Zs\ at 1.86 Myr.  (c)  Initial collapse of the fluctuation to a dense core in $Z =$ 0.04 \Zs\ gas at 3.05 Myr.  (d)  Runaway collapse of the core at 3.063 Myr.}
\label{fig:phase}
\end{figure}

% EDF 4 

\begin{figure}
\centering
\begin{tabular}{cc}
\includegraphics[angle=90, origin=c, width=0.5\textwidth]{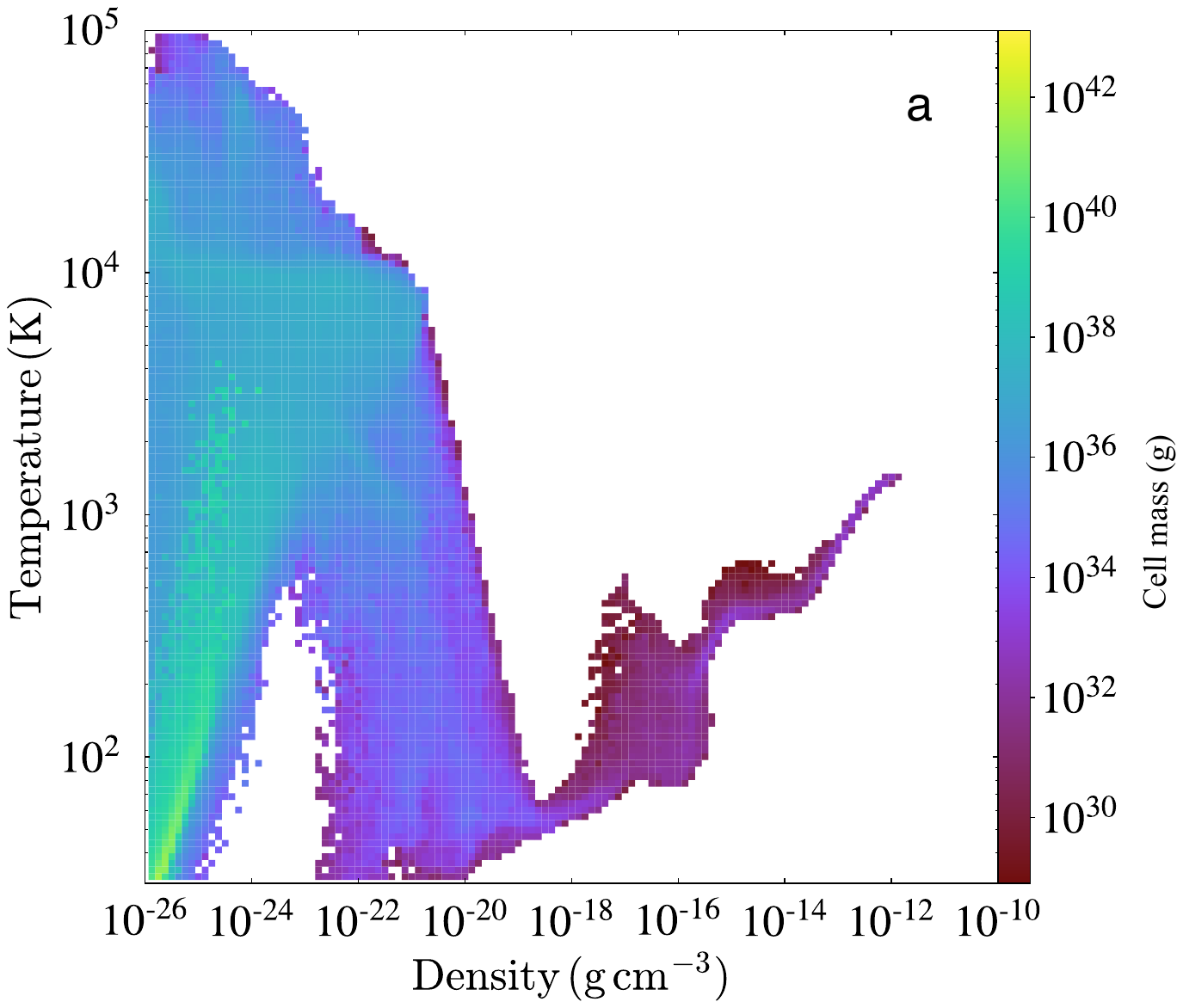} &
\includegraphics[angle=90, origin=c, width=0.5\textwidth]{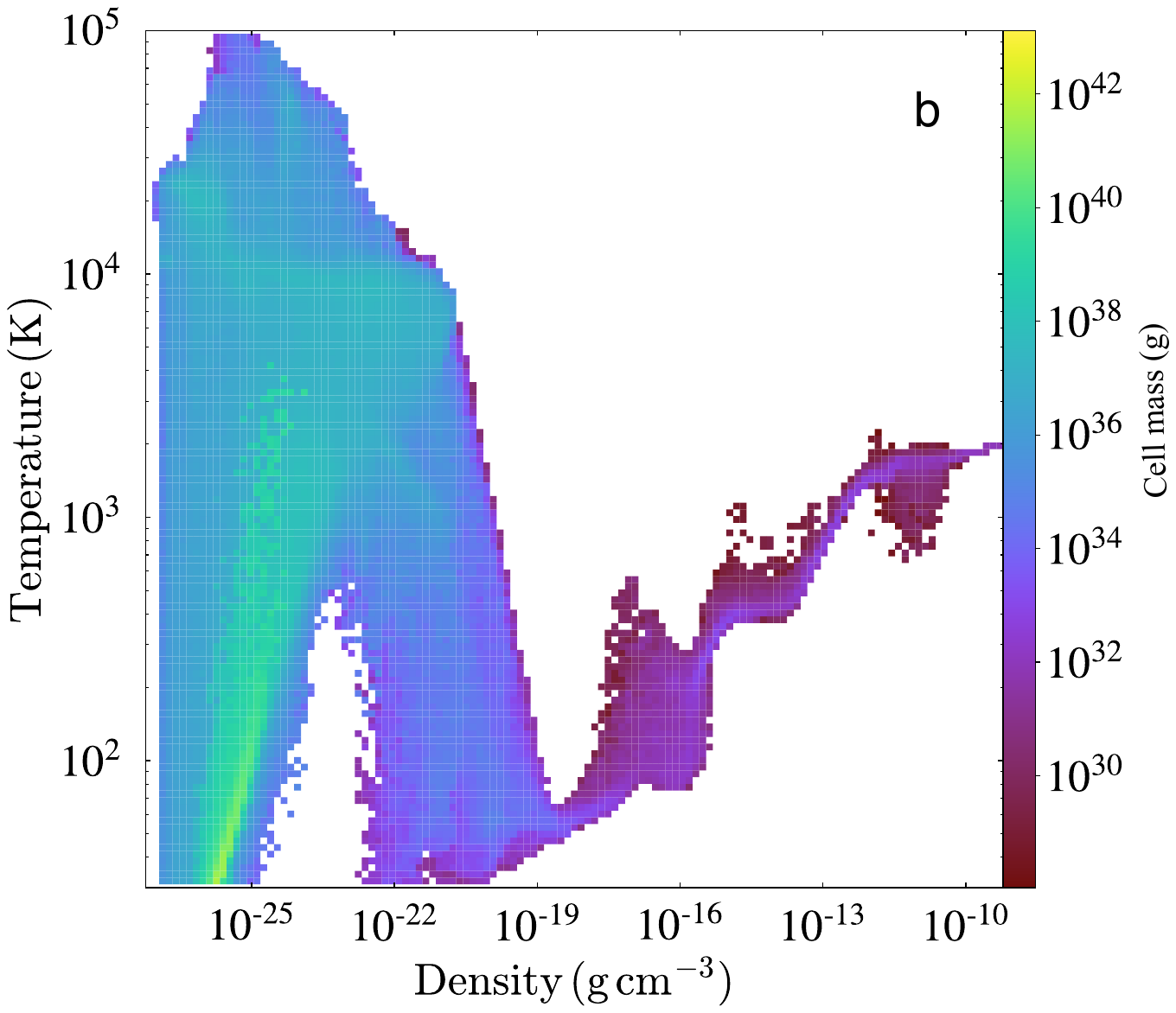} 
\end{tabular}
\caption{{\bf Onset of dust cooling in the PI SN clump.}  As central densities rise from 10$^8$ - 10$^{11}$ cm$^{-3}$, compressional heating increases gas temperatures from 400 - 1500 K as shown in (a).  Dust cooling is then activated, and levels off central temperatures at $\sim$ 1200 K, cooling some of the gas down to 700 K as it collapses to $\sim$ 10$^{14}$ cm$^{-3}$ by 7.7 kyr as shown in (b).  Similar dips in temperature have been found in one-zone collapse tests with dust cooling\cite{om05}.  Not all the gas in the clump falls to a single temperature as in one-zone tests because of hydrodynamical effects in 3D such as turbulence, shocks and compressional heating, but the effects of dust cooling are evident in the dip in temperature in some of the gas at those densities.  At present, we cannot follow the collapse of the dense core in the CC SNR to densities at which dust cooling becomes important.}
\label{fig:dust}
\end{figure}

% EDF 5

\begin{figure*}
\begin{center}
\includegraphics[width=0.45\textwidth]{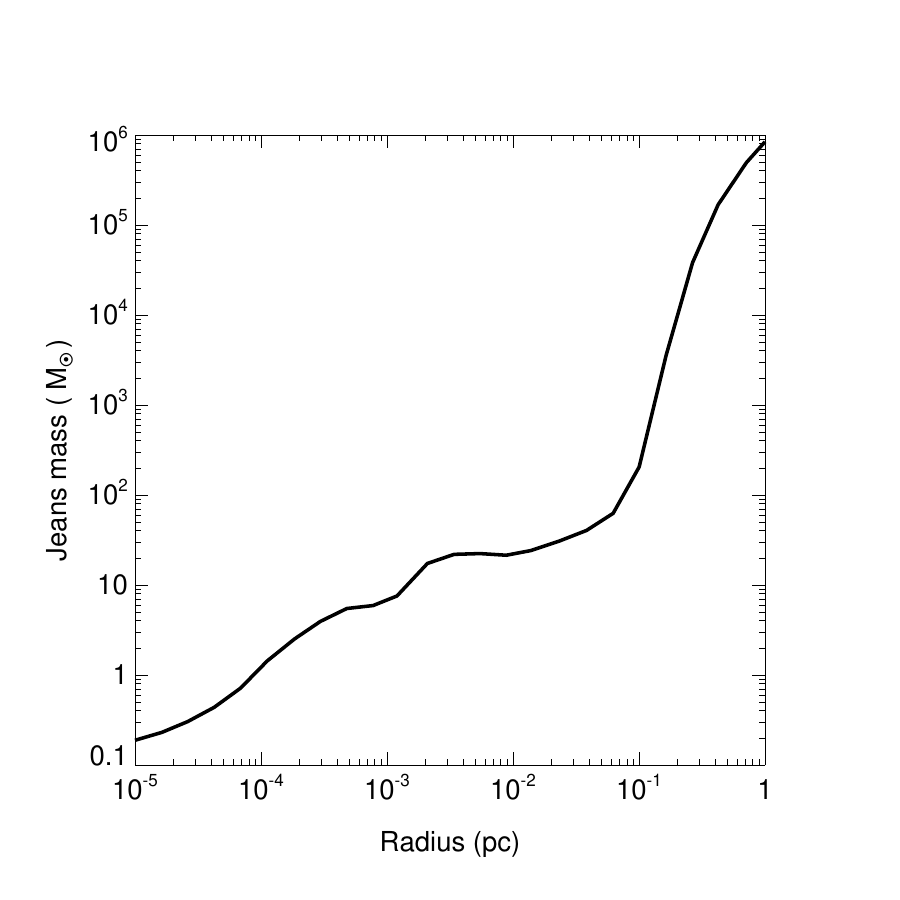} 
\end{center}
\caption{{\bf Jeans masses in the PI SN core.} Gas at the center of the core can fragment on mass scales of $\sim$ 1 \Ms.}
\label{fig:jmass}
\end{figure*}

% EDF 6 

\begin{figure}
\centering
\begin{tabular}{cc}
\includegraphics[angle=90, origin=c, width=0.5\textwidth]{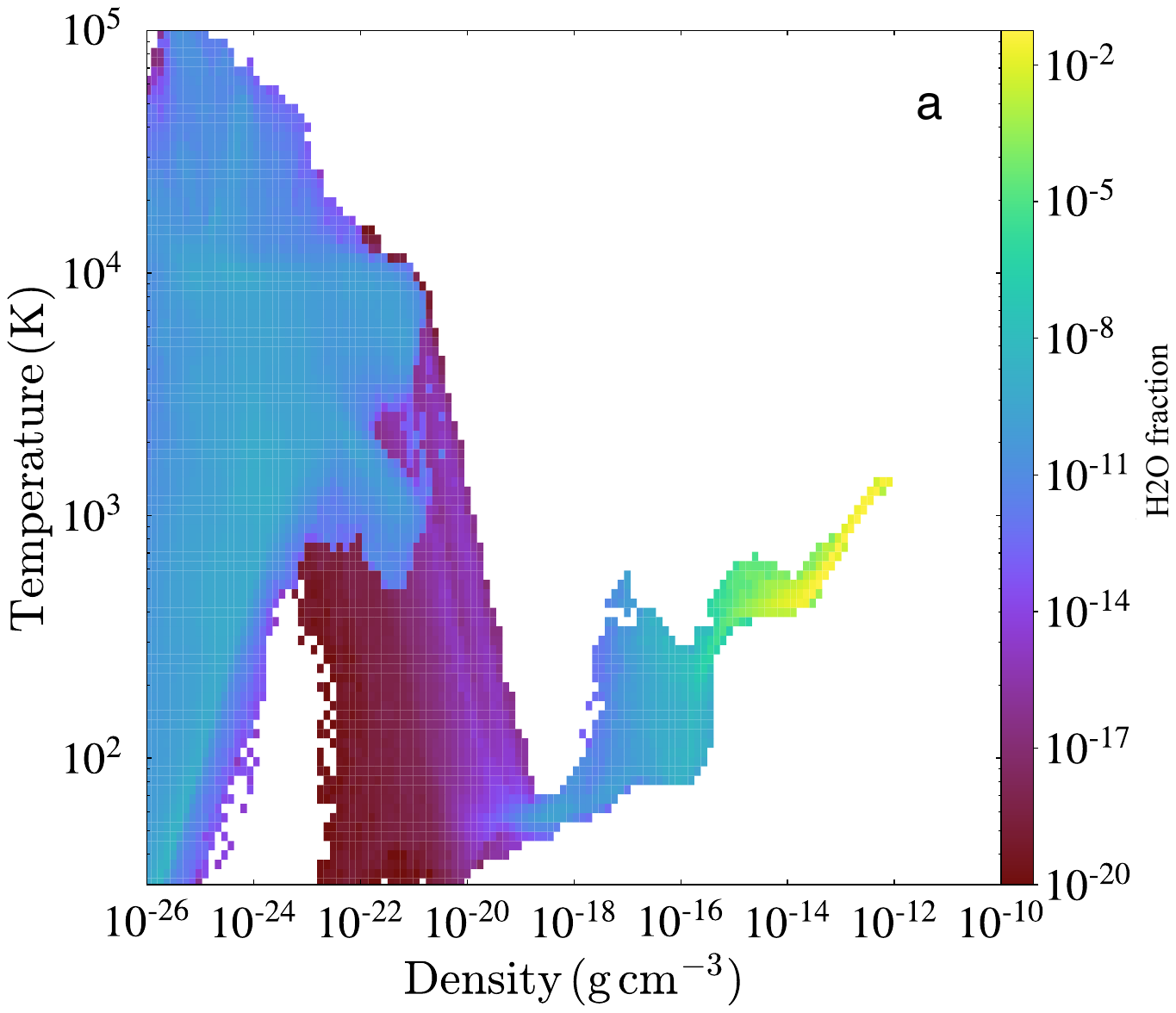} &
\includegraphics[angle=90, origin=c, width=0.5\textwidth]{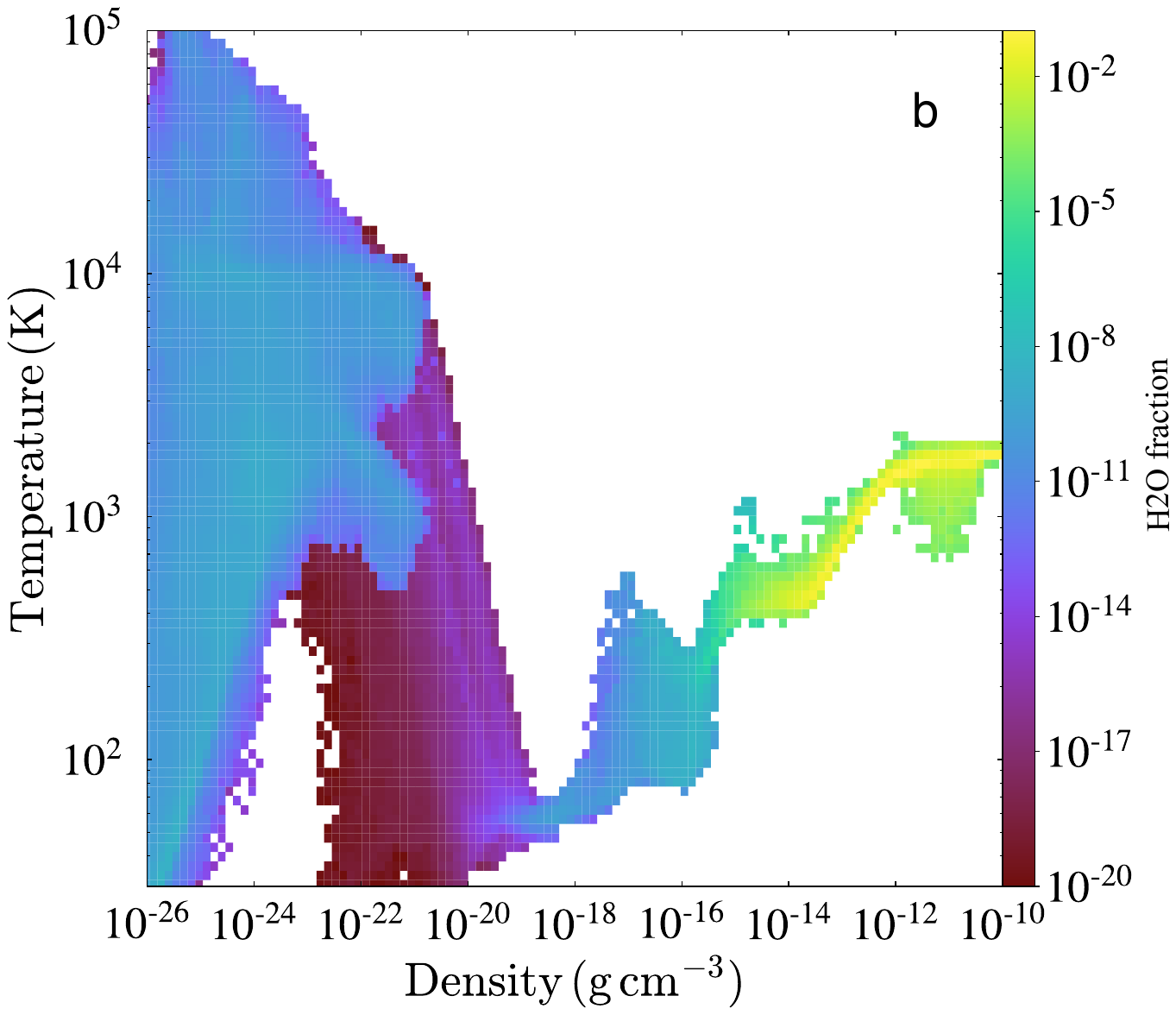} 
\end{tabular}
\caption{{\bf Water formation in the PI SN clump at the times in Extended Data Figure~\ref{fig:dust}.}  Compression and shock heating raise gas temperatures from 400 - 1500 K as central densities rise from 10$^8$ - 10$^{11}$ cm$^{-3}$.  Above $n \sim$ 10$^8$ cm$^{-3}$, three-body formation of H$_2$ rapidly molecularizes the core, which accelerates H$_2$O formation via reactions Z10 - Z12 as shown in (a).  When dust cooling levels central temperatures off at 700 - 1200 K, cooling some of the gas down to 700 K as it collapses to $\sim$ 10$^{14}$ cm$^{-3}$, water reaches mass fractions of 10$^{-4}$ 7.7 kyr later as shown in (b).}
\label{fig:water}
\end{figure}

% EDF 7

\begin{figure*}
\begin{center}
\includegraphics[width=0.5\textwidth]{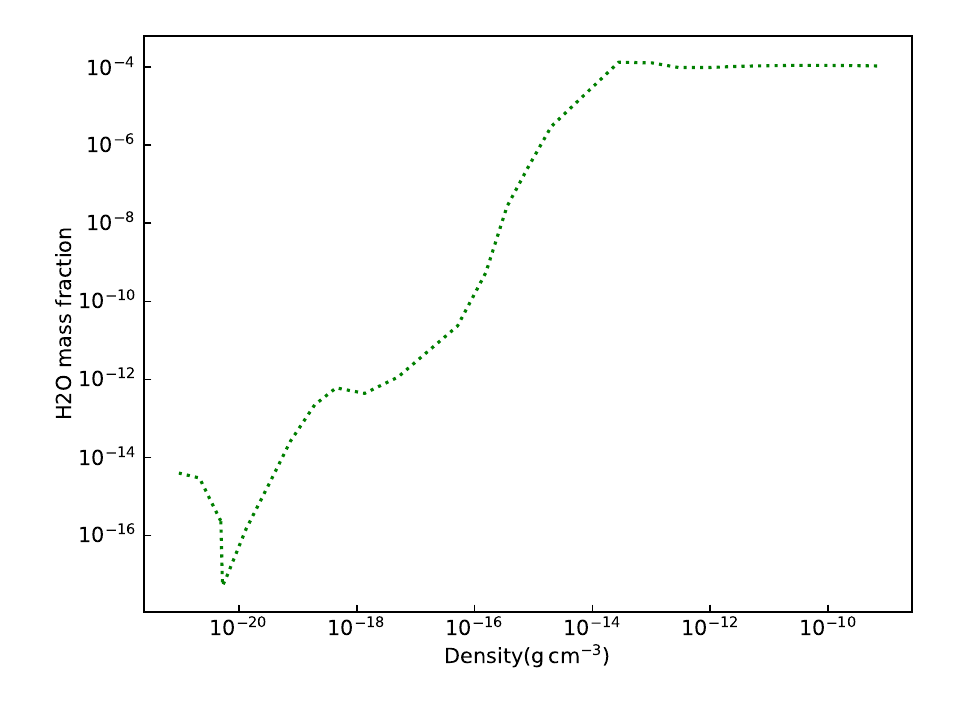} 
\end{center}
\caption{{\bf H$_2$O mass fractions versus central density in the PI SN core.} Central water mass fractions level of at $\sim$ 10$^{-4}$ above $n \sim 10^{10}$, after three-body production of H$_2$ has mostly molecularized the core.}
\label{fig:wmfrac}
\end{figure*}

% EDF 8

\begin{figure*}
\begin{center}
\includegraphics[width=0.5\textwidth]{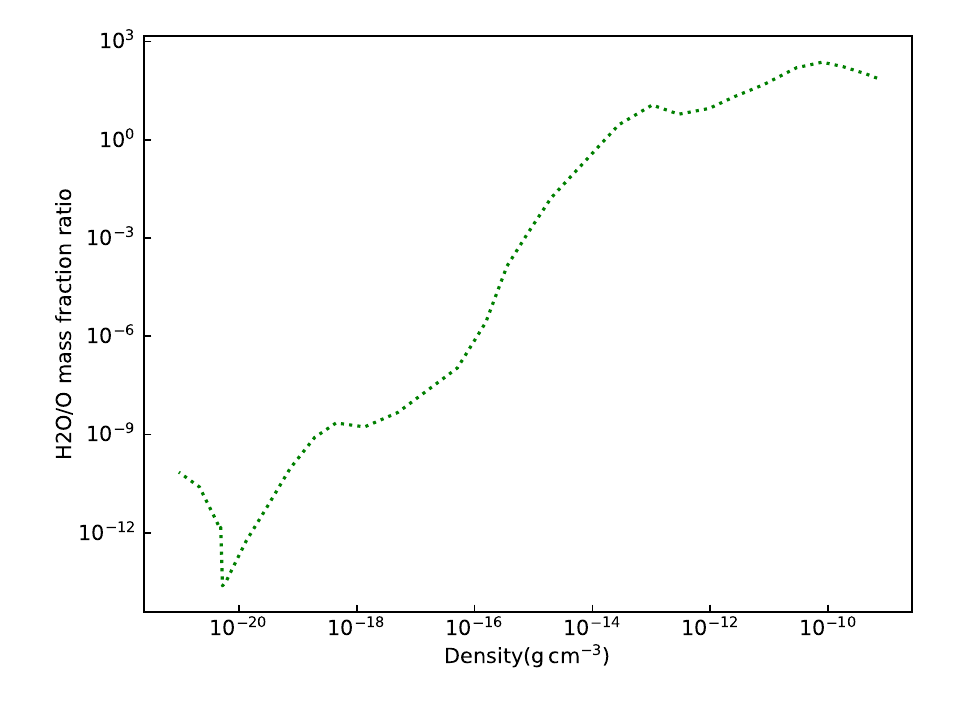} 
\end{center}
\caption{{\bf H$_2$O / O mass fraction ratios in the PI SN core.} Above $n \sim 10^{10}$ when three-body production of H$_2$ begins to fully molecularize the core, H$_2$O / O mass fraction ratios exceed 1 and reach $\sim$ 100, showing that most O depletes to H$_2$O rather than O$_2$.}
\label{fig:h2ofrac}
\end{figure*}

\end{document}